# Birth and death of links control disease spreading in empirical contact networks

Petter Holme[1,2,3,4]

Fredrik Liljeros[3,4]

[1]Department of Energy Science, Sungkyunkwan University, 440-746 Suwon, Korea

[2]IceLab, Department of Physics, Umeå University, 90187 Umeå, Sweden

[3]Department of Sociology, Stockholm University, 10961 Stockholm, Sweden

[4]Institute for Futures Study, 10131 Stockholm, Sweden

E-mail address: holme@skku.edu

**Abstract**

We investigate what structural aspects of a collection of twelve empirical temporal networks of human contacts are important to disease spreading. We scan the entire parameter spaces of the two canonical models of infectious disease epidemiology—the Susceptible-Infectious-Susceptible (SIS) and Susceptible-Infectious-Removed (SIR) models. The results from these simulations are compared to reference data where we eliminate structures in the interevent intervals, the time to the first contact in the data, or the time from the last contact to the end of the sampling. The picture we find is that the birth and death of links, and the total number of contacts over a link, are essential to predict outbreaks. On the other hand, the exact times of contacts between the beginning and end, or the interevent interval distribution, do not matter much. In other words, a simplified picture of these empirical data sets that suffices for epidemiological purposes is that links are born, is active with some intensity, and die.

**Introduction**

To understand, predict and prevent the spreading of infectious diseases in socially structured populations, one needs to know the network over which the disease spread. Over the last two decades, researchers have made much progress by assuming that individuals are connected into networks, and that pathogens can propagate between two individuals connected by a link. This approach, *network epidemiology* (1–3), has changed our view of disease spreading by shifting the focus from properties of populations to the behavior of individuals and their positions in the contact structure of the populations. Even if network epidemiology is a step towards greater realism, it comes with a big simplification—it ignores the timings of contacts. Several studies show that this can lead to big errors in predicting the outbreak dynamics.[4–12] However, it has so far been unclear which temporal structures causes the deviations. In this paper, we try to elucidate this question. To rephrase it, we try to find a good way to simplify empirical temporal networks that keeps as much important temporal and topological information as possible, but not more. We call such a method for simplification a *picture*, and we contrast two similarly complex pictures. One picture—the *ongoing link picture*—is that what we see in a temporal network is the contacts over an unchanging, static network. In this picture, the important temporal structure is the time between events over an edge (*interevent intervals* for short). The other picture—the *link turnover picture*—assumes that the important events are the birth and death of links, or relationships, between individuals. The ongoing link and link turnover pictures are illustrated in Fig. 1.

There has been a lot of effort to understand how interevent intervals affect spreading phenomena.[9–13] The background is that in traditional models of disease spreading one has, explicitly or not, typically assumed homogeneous (Poisson distributed, or periodic) interevent intervals. Empirical data, on the other hand, show interevent intervals that both have fat-tailed distributions ("bursty activity") and imprints of weekly and daily cycles.[14,15] Authors have, by analytical and computational techniques, characterized how such bursty interevent intervals alter the propagation speed and final extent of emerging outbreaks.[16,17] To take this approach to modeling disease spreading on temporal networks is to use the ongoing link picture. Heterogeneous interevent intervals are, however, not the only possible temporal pattern of empirical contact data. If the sampling time was long enough, we would, in most social network, see the birth of death of individuals and the relationships between them. If the birth and death of nodes and links is regarded as the fundamental temporal structure, then we arrive at the link turnover picture. This will make the set of possible transmission routes for a contagious disease to look more like a river system where it is possible for the disease to reach downstream and also a bit upstream, but not far. Studies often describe either one of these aspects—interevent intervals[15,19–21] or birth and death of links[22–25]. We note that some studies[14,26] try to paint an even fuller picture and include statistics about both interevent intervals and the time from the beginning of the sampling time to the first contact between a pair (the *beginning interval*), or from the last contact to the end of the sampling time (the *end interval*). We are not, however, aware of a paper addressing which one of these aspects that is most relevant for disease spreading on empirical data. This is the goal of our paper and we will mostly investigate it by contrasting the two mentioned pictures.

The starting point for our study is twelve temporal network data sets of different forms of human interaction—some arguably relevant for disease spreading, others probably not. Since temporal network data of human behavior is difficult to gather, we use the not-so-relevant data sets too. These are also interesting in a larger context of patterns of human activities,[27] specifically for studies of disease spreading in electronic media.[12,28] After further motivating the study by investigating the statistics of the interevent-, beginning and end intervals for contacts between pairs, we simulate epidemic outbreaks on these data sets. Specifically, we scan the entire parameter spaces of the SIR and SIS models and compare the predictions for the original data sets to modified data where everything is the same as the original, except the feature we are investigating (the heterogeneity of the interevent intervals, the times between the beginning of the sampling and the first contact and between the last contact and the end of the sampling). By this procedure, we can compare the effects of these structures.

## Results

*Empirical networks*

We analyze human contact sequences of different kinds (all empirical data sets of this sort that we have access to). The data sets can be divided into two classes—electronic communication and physical proximity. The latter class is more relevant for epidemiological purposes. In the online communication class, we study two e-mail networks—from refs. 29 (*E-mail 1*) and 30 (*E-mail 2*). Even though an e-mail is naturally directed, to analyze all the data in the same way, we treat e-mails as undirected contacts. The e-mail data sets are sampled from a set of e-mail accounts. A difference between them is that *E-mail 1* includes contacts to external e-mail accounts while *E-mail 2* only records e-mails between the sampled accounts. This is a kind of sampling boundary effect that can be avoided by studying communication within a closed community. We do this with data sets from three Internet communities: two dating communities—*Dating 1*[26] and *Dating 2*[31]—and a film-rating community[32] (*Film*). A different form of online communication is posts to public web pages. We study one data set of posts to the home page ("wall") of other users at the social network service *Facebook* (data from ref. 33) and a data set from the above-mentioned cineaste community where a contact represents a reply to a post at a public forum[32] (*Forum*). The first physical proximity data set we use comes from the *Reality Mining* study[34] where students were equipped with smartphones, and a contact were recorded if they were in Bluetooth range (about 10 m) from each other. We use the same subset of this data as ref. 35—a one-week subsample of the full data. It was cleaned and discretized to one sample per five minutes. Another type of proximity data was gathered by radio-frequency identification sensors worn by the participants of a conference[36,37] (*Conference*) and visitors to a gallery[36,37] (*Gallery*). The latter is split into 69 distinct days, which we will analyze separately (rather than combining the to one contact sequence). In these data sets, a contact is recorded at 20 seconds intervals between individuals within a range of 1–1.5m. Another data set of proximity comes from a regional health-care system where people are recorded as in contact if they are at the same ward at the same time (*Hospital*).[38] Finally, we use a data set of sexual contacts between escorts and sex buyers collected from web forum (*Prostitution*).[39] We list some basic statistics of the data sets in Table 1.

*Preliminaries*

The data sets we discuss in this paper are sequences of *contacts*—triples $(i,j,\tau)$ recording that individuals $i$ and $j$ have been in contact at the (discrete) time $\tau$.[14] The sampling starts at time 0 and ends at time $T$. For a specific pair of individuals $(i,j)$, we talk about the set of times $T_1 < T_2 < \ldots < T_l$ of contacts involving $i$ and $j$ as their *contact history*. The time between the beginning of the data and the first contact, $t_B = T_1$, is the *beginning interval* of the contact history. Similarly, we call the time between the last contact to the end of the sampling time—$t_E = T - T_l$—the *end interval*. The times between the contacts—$t_i = T_{i+1} - T_i$—are the *interevent intervals* of the link. These concepts are illustrated in Fig. 2.

*The ongoing link picture*

The initial motivation for our work is that many empirical temporal-network data sets (indeed all data sets available to us) are statistically inconsistent with a picture of a network of individuals connected throughout the sampling time. Refs. 14 and 25 also note that empirical contact data are inconsistent with this assumption. More specifically, for many of the links, the beginning and end intervals expected, given the interevent intervals, are much shorter than the observed values. A parsimonious description consistent with these observations is that links are born and die during the sampling. In Fig. 3, we plot the fraction $\phi$ of links where $t_B$ and $t_E$ and that are significantly different ($p < 0.05$) from their values predicted from the interevent intervals according to the method described in the Methods section. Briefly, one needs observe that (assuming the ongoing-link picture is correct) the probability there will be a long interval at the beginning of the sampling is longer simply because the interval is longer, there is also a penalty for long intervals since they need to finish before the sampling time is over, they have a shorter period they can begin. These two effects do not cancel, so that one arrives at a somewhat complicated formula for calculating the expected distribution of $t_B$ and $t_E$. If the ongoing link picture holds perfectly, this fraction would thus be 0.95, but it is at most a little over 0.5 (for the *E-mail 2*) data.

On a side note, we see that for many (especially electronic communication) data sets (in Fig. 3), $\phi$ is larger for the end intervals than beginning intervals. This means that more links are active at the end compared to the beginning, which is natural in the cases the data represents a growing community of individuals (which is the case for the *Dating 1*, *Forum*, *Film* and *Prostitution* data).

We can conclude that, from a statistical point of view, the real data does not fit the ongoing link picture. This observation alone does not disqualify it as a picture of empirical contact sequences—it could be these statistical properties are irrelevant for disease spreading. Unfortunately there are no equally straightforward predictions to test for the link turnover picture. To compare the two pictures, we therefore need to simulate disease-spreading models on the empirical contact sequences and reference data where we can control the effects of the beginning-, end- and interevent intervals.

*The link turnover picture and reference data*

In the link turnover picture, links are born, are active for a while, with a certain number of contacts, and die. To argue for this picture, we contrast the behavior of the SIR and SIS models on empirical temporal networks to the behavior on reference data were we modify the beginning-, end- and interevent intervals. If we e.g. destroy the structure of the interevent intervals, but not the beginning- and end intervals, and that changes the result much, then we conclude that interevent intervals significantly affect disease spreading. In particular, to test the effect of the interevent intervals (without changing the beginning- or end intervals), we make them equal (replacing $\tau_1, \tau_2, \ldots, \tau_L$ by $\tau_1, \tau_1 + \Delta\tau, \tau_1 + 2\Delta\tau, \ldots, \tau_L$, where $\Delta\tau = (\tau_L - \tau_1) / (L - 1)$). ($L$ is the number of contacts across the link.) This changes the interevent interval distribution from a (usually) heterogeneous to a degenerate distribution of only one possible interevent interval. We call this type of reference data *Interevent intervals neutralized* (IIN). To test the effects of the beginning- and end intervals (without changing the interevent intervals), we shift the contact histories of the pairs so that they all start at the beginning of the sampling time (replacing $\tau_1, \tau_2, \ldots, \tau_L$ by $0, \tau_2 - \tau_1, \ldots, \tau_L - \tau_1$). Similarly, we also shift the links to all end at the same time (replacing $\tau_1, \tau_2, \ldots, \tau_L$ by $\tau_1 + T - \tau_L, \tau_2 + T - \tau_L, \ldots, T$). We call these reference data *Beginning intervals neutralized* (BIN) and *End intervals neutralized* (EIN) respectively. The methods to construct the reference data are illustrated in Fig. 4. Note that the (static) networks of accumulated contacts remain the same for all types of reference data.

One can think of other ways of altering the beginning-, end- and interevent intervals for the same purpose. For example, one could generate new beginning intervals from the interevent intervals, similar to the analysis above (in the context of Fig. 3). Such randomization schemes are otherwise very useful for temporal networks.[17] The main reason not to chose this path has to do with our computational

constraints—the largest data sets took us over a core year per parameter value to simulate, so averaging over an ensemble model would be quite intractable.

*SIR model*

We run the SIR model on the empirical and reference temporal network data and focus on the final fraction of individuals infected as some time, $\Omega$. We scan the entire parameter space—the per-contact transmission probability $\lambda$ and the disease duration $\delta$ (measured in units of *T*). For more details on the simulation, see the *Methods* section. In Fig. 5, we plot a typical output (in this case for the *Prostitution* data). Comparing the simulation based on empirical data to that based on the IIN reference data, the $\Omega$ values look strikingly similar. The BIN and EIN reference data give rather different $\Omega$ values, especially for high transmission probabilities and low durations of the infectious state. This has an intuitive explanation in that both the BIN and EIN reference data shorten the effective sampling duration, so the infectious period does not have to reach that far for the disease to spread between two individuals. See Supplementary Information S1 for the plots corresponding to Fig. 5 for the other data sets. (In this file, one would have to flip through the PDF pages to see the difference between the panels corresponding do different reference data.)

We summarize the SIR simulations by measuring the average deviation of $\Omega$ between the empirical and reference data, $\overline{|\Delta|}$. These are displayed in Fig. 6. The main picture is clear. Destroying the interevent interval distribution does not change $\Omega$ as much, on average, as destroying effects of the beginning- and end intervals; and this is true for all data sets. Looking at the extreme, rather than the average, deviations gives the same conclusions (even though the differences are not as big)—see Supplementary Information S2. These results are in favor of the link turnover picture. Just like the analysis of the predictability of beginning- and end interval distributions above, the *Dating 1*, *Facebook* and *Conference* data shows comparatively small deviations (and thus fitting less bad to the ongoing link picture than the other data sets). At the same time, $\phi$ does not predict $\overline{|\Delta|}$ perfectly—e.g. *Facebook* has larger $\phi$ values than *Prostitution*, but also larger $\overline{|\Delta|}$ values.

*SIS model*

As it allows reinfection, the SIS model is a bit more complex to analyze than the SIR model. We calculate both the average fraction of individuals at one point infected and the average number of infections per individual. We use the same parameter values as the SIR analysis. Even though the numbers are different, the conclusion is qualitatively the same as for the SIR model. Shifting the links to the beginning and end of the data affect the disease spreading much more than making the interevent intervals equal. See Fig. 7 for the $\overline{|\Delta|}$ results for the SIS model. In Supplementary Information S3, we show the plots corresponding to S1 for the SIS model; in S4 we plot the average number of infections per individual for all empirical and reference data; in S5 we plot the maximal deviations of the average outbreak sizes.

Note also that Fig. 5, 6 and 7 confirms the basic assumption behind this work—that there are temporal structures that cannot be simplified by reducing the data to a static network without losing information relevant to the disease spreading. In static network epidemiology, all panels of Fig. 5 would be the same and all bars of Figs. 6 and 7 would be zero.

**Discussion**

To fully characterize the contact patterns of a population for the purpose of understanding disease spreading, one needs to know who that has been in contact with whom at what time. Although such sequences of contacts are mathematically rather simple—just pairs of individuals and times of their contacts—they are conceptually difficult. For example, drawing a contact sequence rarely leads to some intuition about its effect on disease spreading, not even for a small data set. This calls for a further simplification of contact data, which still captures epidemiologically relevant temporal structures. Our results tells us that, for the purpose of infectious disease spreading, it makes sense to think of our empirical data as links—pairs of people in contact at some time—with a certain beginning and end, and number of contacts. We contrast this to a picture of links being active before the beginning and after the end of the sampling. It is easy to see that the latter, ongoing link picture is statistically inconsistent with our empirical data (in the sense that the beginning and end-intervals are too long to result from the same process as the interevent interval distribution). This alone does not disqualify the ongoing link picture for the purpose of studying disease spreading—it could be that the links consistent with the ongoing link picture are the ones that matter for disease spreading. To argue further that the link turnover picture is appropriate for disease spreading on our empirical data sets, while the ongoing link picture is not, we compared the SIR and SIS models on the original data to three types of reference data. If we destroy the interevent intervals by making them all equal, then there is not so much of a difference in outbreak sizes. This means that the interevent interval distributions—the fundamental temporal structures in the ongoing link picture—are not important for the outbreak sizes. What matter more, on the other hand, are the beginning and end intervals. If these are destroyed by letting all links begin or end simultaneously, the effect can be very strong. For example, for the *E-mail 1* data at parameter values $\delta = 0.01$ and $\lambda = 1$ the original data has an average outbreak size of 0.1% of the population (around three individuals) while the BIN reference network has outbreaks averaging 79% of the population.

There have not been many studies trying to sort out which structural aspects of the contact sequences that are important for disease spreading. Studies typically either assumes the ongoing link[9–12] or link turnover picture.[5,39] Ref. 40 has somewhat similar goals to ours as they investigate, for some proximity data sets, which temporal-network structures that are important for spreading in a SI model with 100% transmission probability (the SIR or SIS models with $\lambda = \delta = 1$). The authors conclude that (which is compatible to our results) the distributions of links per individual and contacts per links are enough to model the spreading. Furthermore, the SIR and SIS models get more sensitive to the contact patterns the smaller $\delta$ is, a model that works for $\lambda = \delta = 1$ might not do so for other parameter values. This also applies to refs. 6, 9, 11 and 12 that use the same spreading model.

Mathematical epidemiology has plenty of concepts to explain how contact structures affect epidemics. In this paper, we found that a new such concept—the link turnover picture—explains the effects of empirical temporal-network structure with respect to disease spreading. This picture relates to other concepts and we will turn to a brief discussion of these connections. First, there is an idea that high concurrency (roughly, the relative amount of simultaneously active links in a population) facilitates disease spreading.[41,42] If we shift the links to begin (or end) at the same time, then the concurrency will increase to its theoretical maximum at the beginning (or end) of the sampling. Other times the concurrency will be lower than the empirical networks. During the peak concurrency, we can expect disease to spread fast and easily (even for relatively low $\lambda$-values). This can be seen in Fig. 5b and c where the outbreak sizes are much larger than in the original data (Fig. 4a). The BIN (EIN) reference data also make the overall contact rates much more concentrated to the beginning (or end), which another (perhaps more fundamental) aspect explaining the larger outbreak sizes for short duration of infections and high transmission probabilities.

Another way of understanding these results is to compare the time-scales of outbreak dynamics and the sampling time. If the sampling time is short enough, the links should either consist of only one contact, or fit the link turnover picture. If the sampling time is long enough, say longer than the lifetime of individuals, then the beginning and end interval of most links should really be their first and last contacts respectively. The contact patterns for real disease spreading are, most likely, somewhere between these extremes. Ultimately, the conclusions of this paper holds for the data sets we study. They can probably be generalized to many, but probably not all, relevant contact patterns. Nevertheless, in the light of our results, we hope to see future research on how the birth and death of links affect disease spreading.

**Methods**

*Predicted cumulative distribution of beginning and end intervals*

To investigate if the beginning- and end intervals can be explained by the interevent interval distributions, we assume that the interevent intervals of a link ($i,j$) are uncorrelated samples from the same distribution $p_{(i,j)}(t)$. We will drop the subscript ($i,j$)—all calculations are valid for any particular link. To get the predicted cumulative distribution of beginning or end intervals $P_{BE}(t)$ (the calculation for end intervals is analogous) from the *observed* interevent interval distribution $p'(t)$, we need three ingredients: First, the probability of observing the interval $t$ is proportional to $T + 1 - t$, i.e. a time window that would allow the time interval to end before the end of the sampling. Second, the probability of observing an interval $t$ as the beginning interval is proportional to $2t$ (the factor $t$ from that longer intervals are more likely to be observed; the factor two from that the observed interval would be, on average, half as long as the real interval). I.e., $p_{BE}(t) \sim t\, p(t) / 2$. Third, the cumulative distribution $P_{BE}(t)$—the chance of seeing an interevent interval larger than, or equal to, $t$—can be obtained from $p_{BE}(\tau)$ by summing all $\tau$-values from $t$ to infinity and dividing its theoretical maximum (the sum of all $\tau$-values). The first two points tells us that $p_{BE}(t) \sim t\, p'(t) / 2(T + 1 - t)$. Also using the

third points, we obtain

$$P_{BE}(t) = \sum_{\tau=t}^{\infty} \frac{\tau p(\tau)}{T+1-\tau} \bigg/ \left[\sum_{\tau=1}^{\infty} \frac{\tau p(\tau)}{T+1-\tau}\right] = \sum_{i:t_i \geq t} \frac{t_i}{T+1-t_i} \bigg/ \left[\sum_i \frac{t_i}{T+1-t_i}\right]$$

(1)

For each link, we measure $P(t_B)$ and $P(t_E)$ and count the fraction of links ϕ where $P(t_B) > 0.05$ or $P(t_E) > 0.05$.

*SIR and SIS simulations*

We simulate the SIR and SIS models by averaging over each individual as a starting point of the infection 1000 times (so in total we perform 1000$N$ simulation runs for each set of parameter values). The initially infectious individual is assumed to be infectious at the time of its first contact. An infected individual stays infectious for a time δ. This is different from traditional, differential-equation based modeling where the period of infectivity is exponentially distributed (which is both further from reality[43], harder to implement in simulations and yields qualitatively similar results[44]). For each data set, we scan 400 coordinates in the δ×λ-space (where λ is the per-SI-contact transmission probability). We use an exponential sequence of 20 steps in both the δ and λ dimension that goes from 0.001 to 1.

**Acknowledgments**

This research was supported by the Basic Science Research Program through the National Research Foundation of Korea (NRF) funded by the Ministry of Education (2013R1A1A2011947) and the Swedish Research Council. The computer simulations were carried out at the Abisko cluster of HPC2N, Umeå University.

**Contributions**

P.H. and F.L. conceived the study; P.H. developed the methods and analyzed the data; P.H. and F.L. wrote the paper. All authors read an approved the final version of the manuscript.

**Competing financial interests**

The authors declare no competing financial interests.

**Corresponding author**

Correspondence to: Petter Holme

**Figure captions**

**Fig. 1. Illustrations of the ongoing link and link turnover pictures.** The network on the top illustrates that the panels below refer to two links in the network (i.e., the contacts between two individuals). In panel a, we see the times of contacts over the two links. Panel b illustrates the ongoing link picture where one assumes the links are continuously active, so contacts could happen outside of the sampling time frame. The interevent intervals in a are the same as in Panel b, but ordered differently. Panel c illustrates the link turnover picture where one considers a link active between the first and last contact, and do not care about the exact timing of intermediate contacts.

**Fig. 2. Illustrations of time related notations.** The network at the top illustrates that the panel below concerns one link. Panel a shows a time line of the contacts between two individuals in a link. The time between the start of the sampling and the first contact is the beginning interval $t_B$; the time from the last contact to the end of the sampling is the end interval $t_E$. Uppercase variable names refer to the time of event, while lowercase refer to the duration of intervals.

**Fig. 3. The fraction of links consistent with the ongoing link picture.** A link is said to be consistent with this picture if the hypothesis that the beginning (or end) intervals is drawn from the same distribution as the interevent intervals has $p > 0.05$. The horizontal line corresponds to the theoretical value $\phi = 0.95$ expected if the data is fully consistent with the ongoing link picture.

**Fig. 4. Illustration of the construction of reference model networks.** Like in Fig. 1, the network above is a reminder that the time lines to the right is for only one of the links. Panel a shows the time line of contacts of the highlighted link to the left. Panel b shows the same link where the interevent intervals are redistributed to the same value, thus removing the heterogeneous interevent intervals while keeping the number of contacts and the beginning and end intervals constant—the IIN reference data. Panel c shows the BIN reference data—contacts of the link with the contact sequence shifted so that the beginning interval coincides with the beginning of the sampling time. Thus the influence of the times to the first contact is destroyed. Panel d shows the dual manipulation to c where the times are moved to the end of the sampling time.

**Fig. 5. The average outbreak size for SIR model, an example.** We use the *Prostitution* data and plot the fraction of at-some-point-infected individuals as a function of the duration of the infectious state $\delta$ and the per-contact infection probability $\lambda$ (for a contact between an infectious and susceptible individual). Panel a shows the values for the original network; b is for the reference network with interevent intervals of equal length (the IIN reference network); c shows values for the BIN reference network (with the first contact of each link happening simultaneously); d shows values for the EIN reference network (with the last contact of each link happening simultaneously).

**Fig. 6. Deviation from the empirical outbreak sizes for the different reference data (SIR model).** This plot shows $|\Omega_{empirical}(\lambda,\delta) - \Omega_{reference}(\lambda,\delta)|$ averaged over $\lambda$ and $\delta$. A large value of this quantity means that the structure that the reference model manipulated is important for disease spreading at that data set. The *Gallery* data are averaged over all 68 distinct data sets, the horizontal bar and shaded areas indicate the average and standard errors respectively.

**Fig. 7. Deviation from the empirical outbreak sizes for the different reference data (SIS model).** This figure is exactly corresponding to Fig. 6 except it is for the SIS model.

**Table caption**

**Table 1.** Statistics of the data sets. We display the number of individuals, contacts, the total sampling time and the time resolution (shortest time between two contacts).

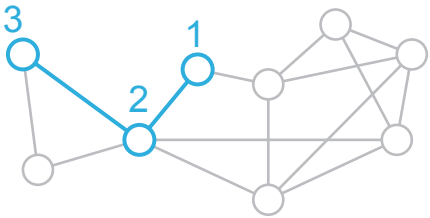

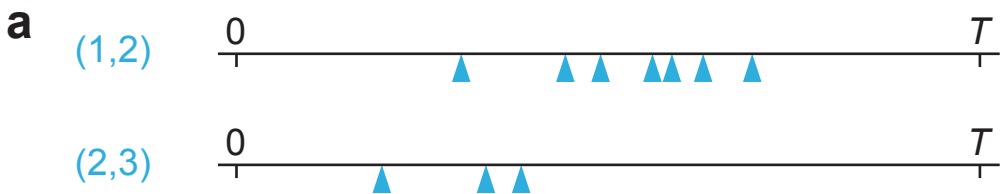

ongoing link picture

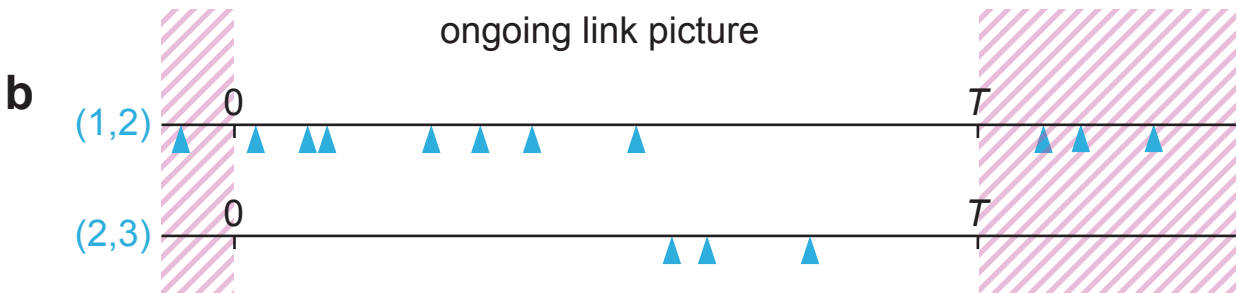

link turnover picture

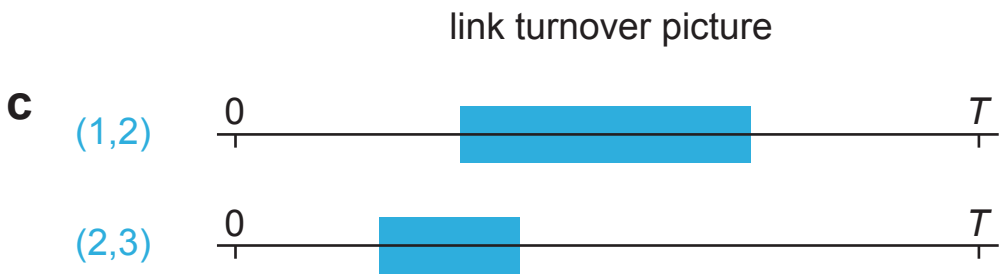

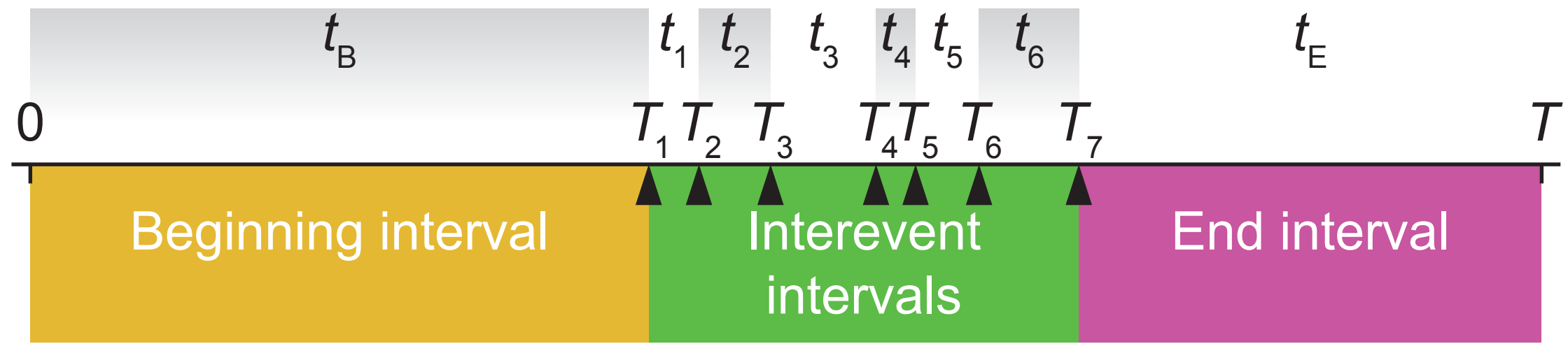

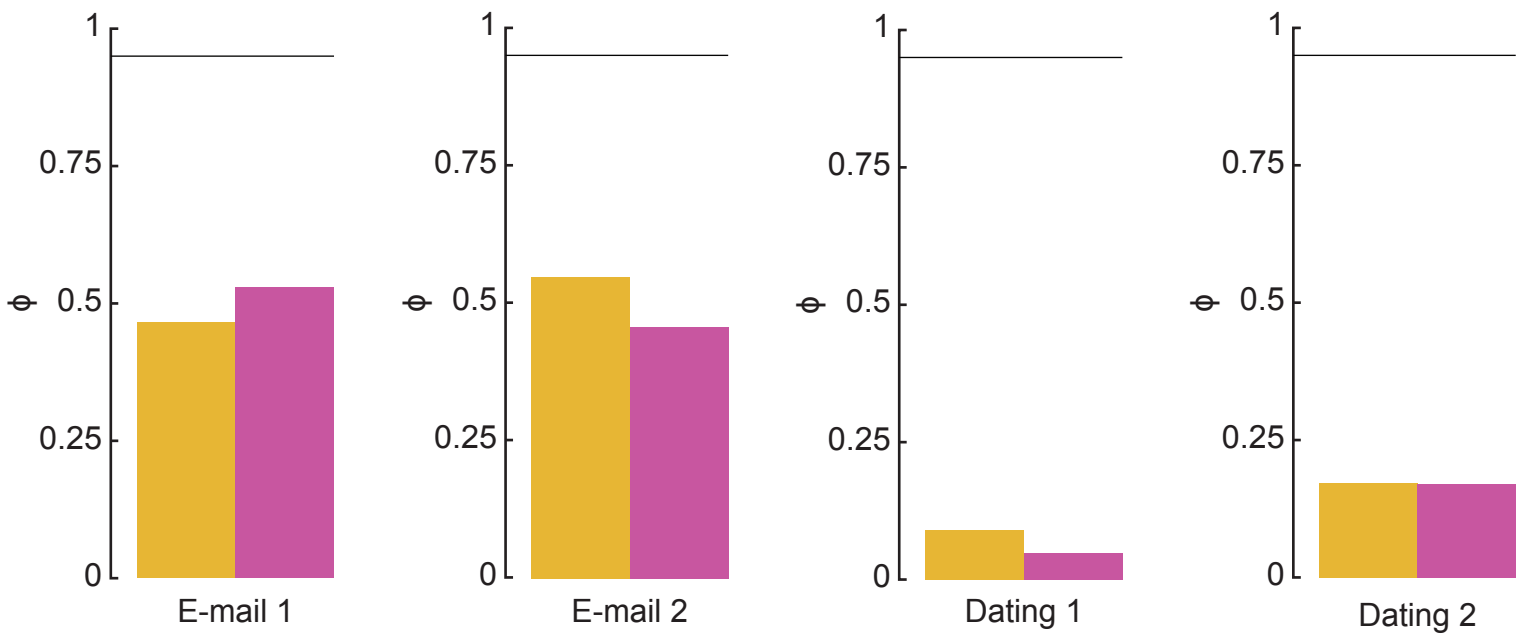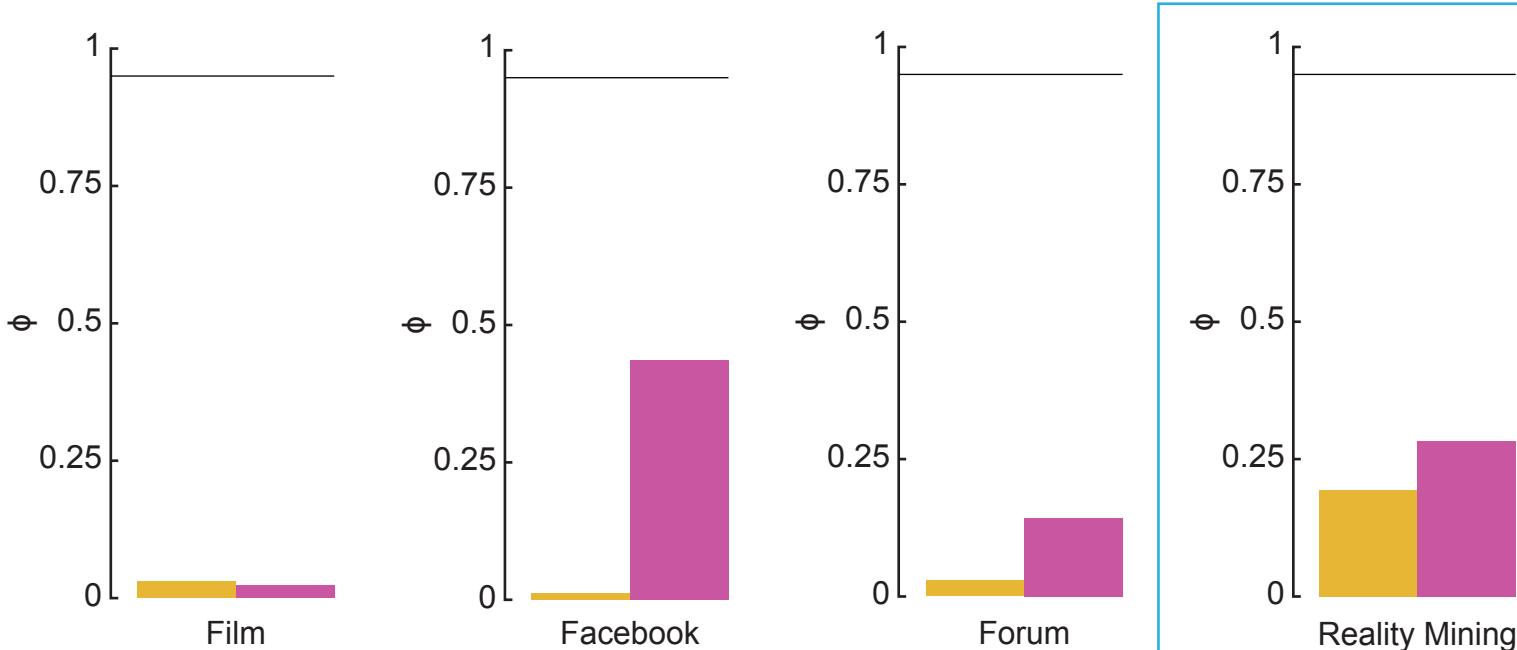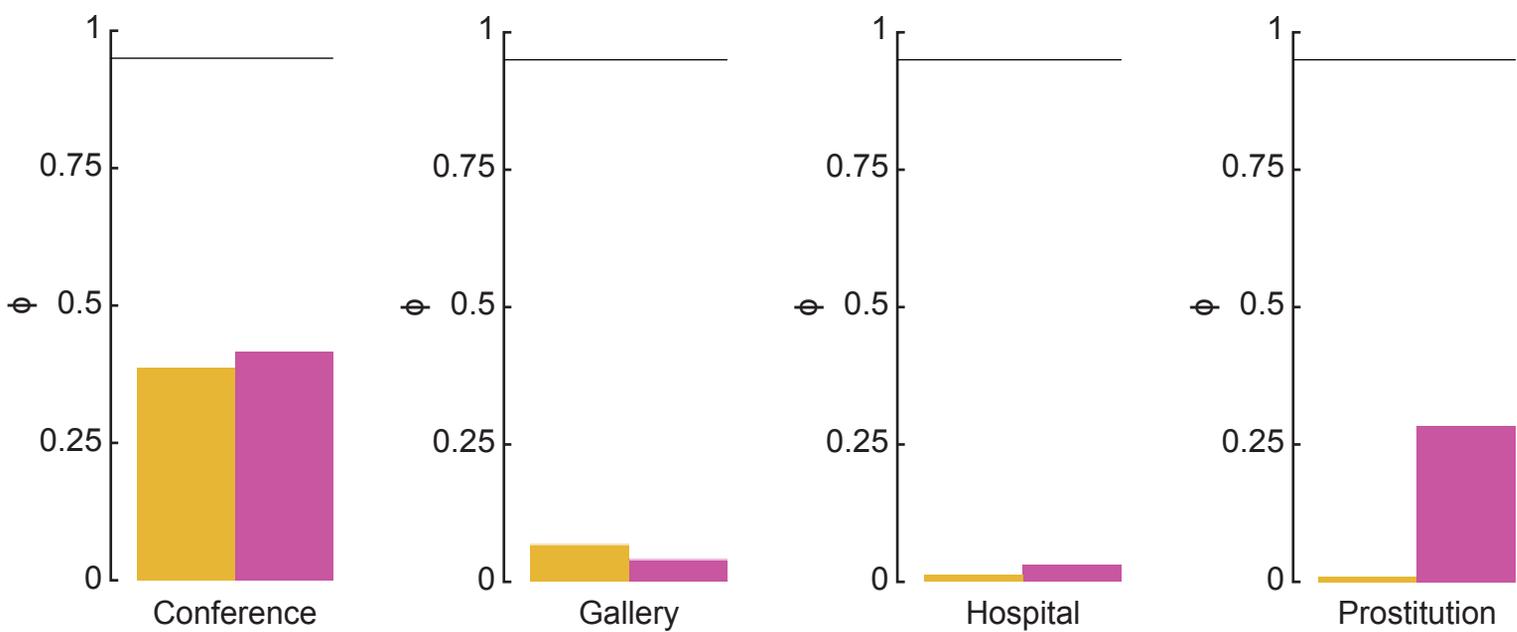

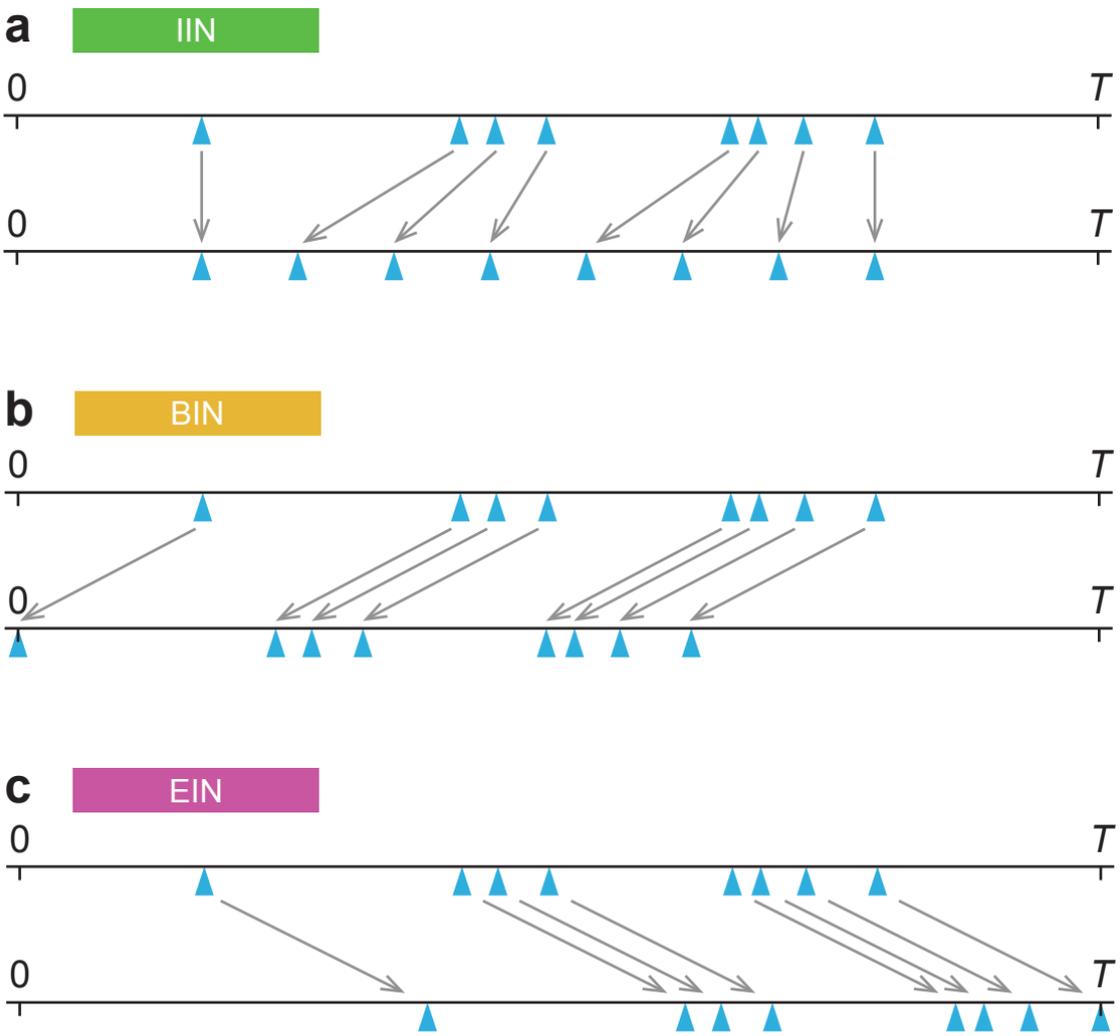

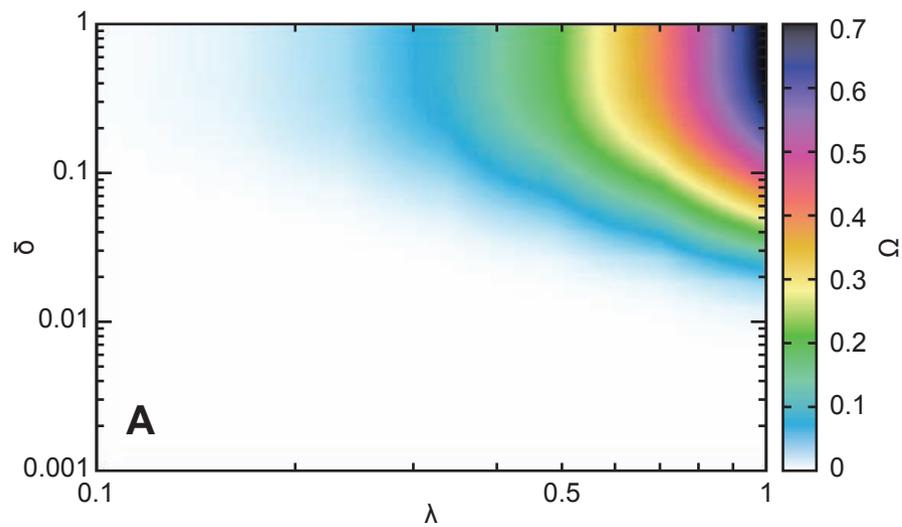
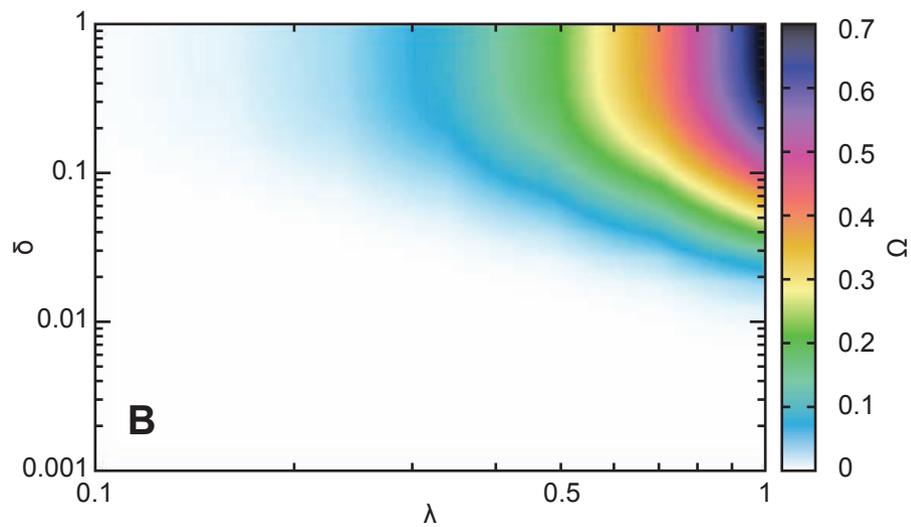
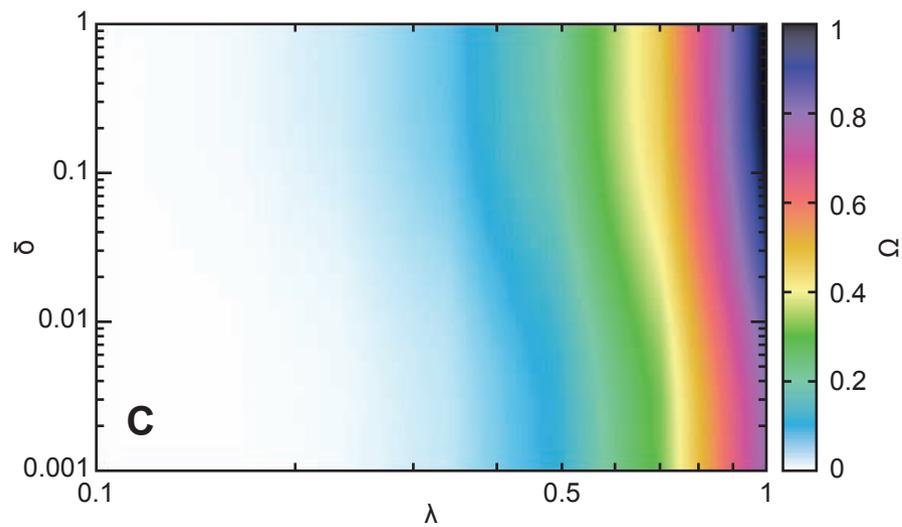
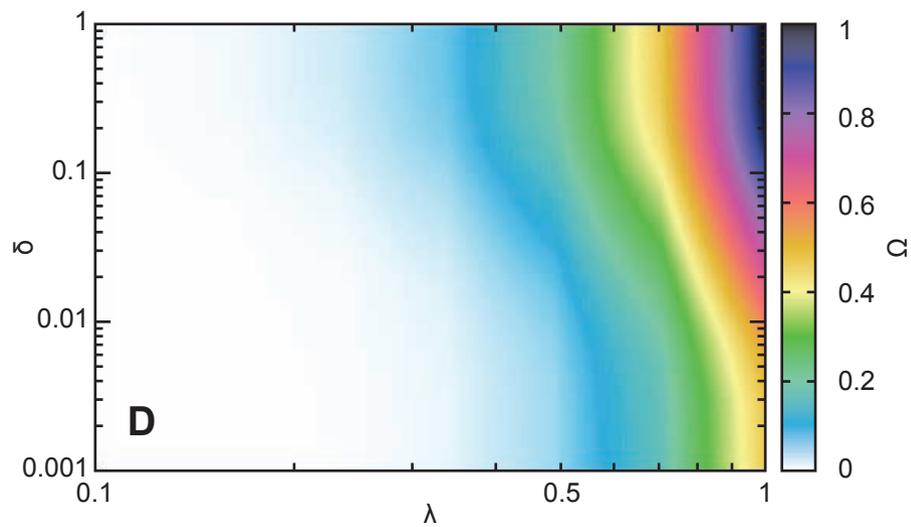

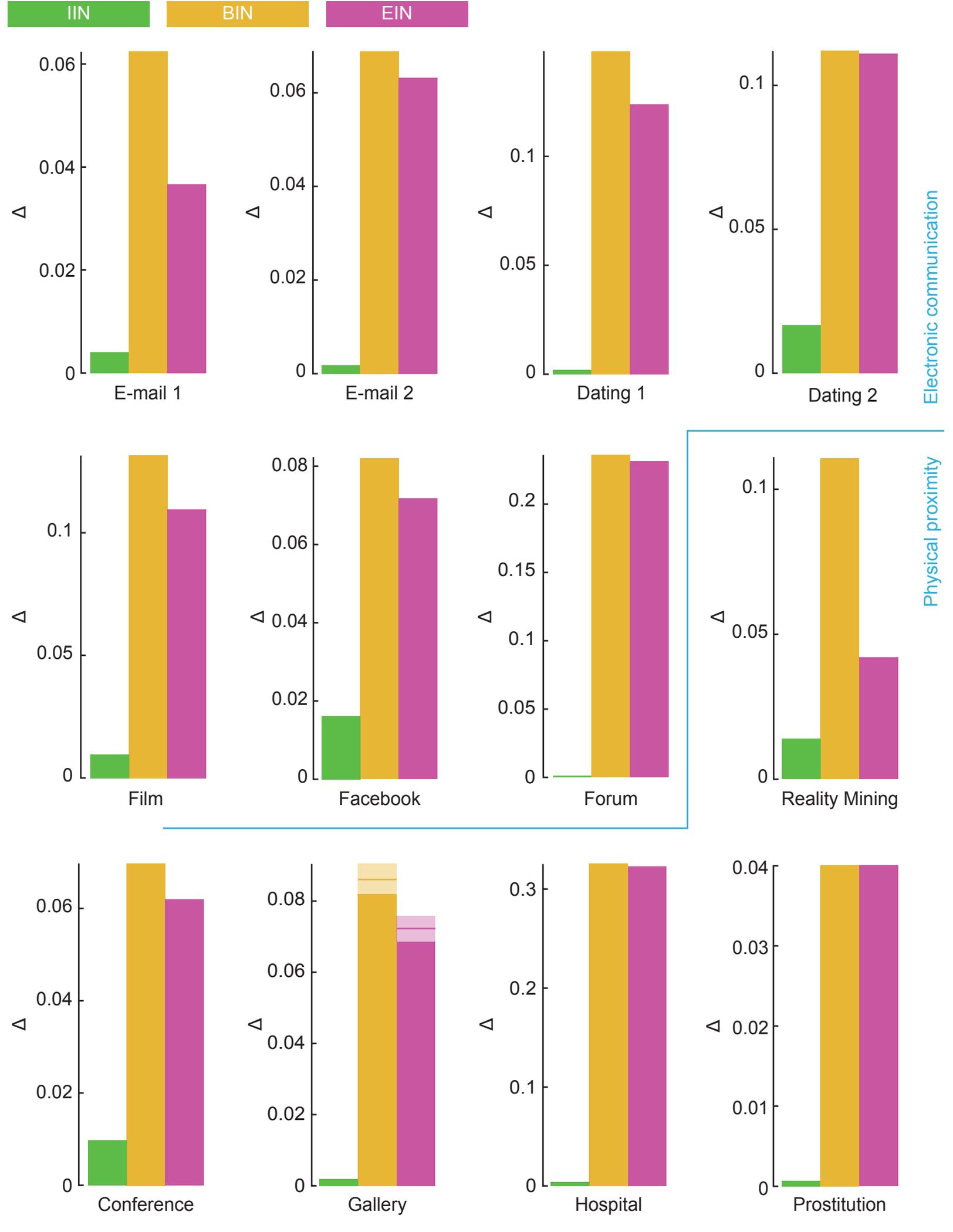

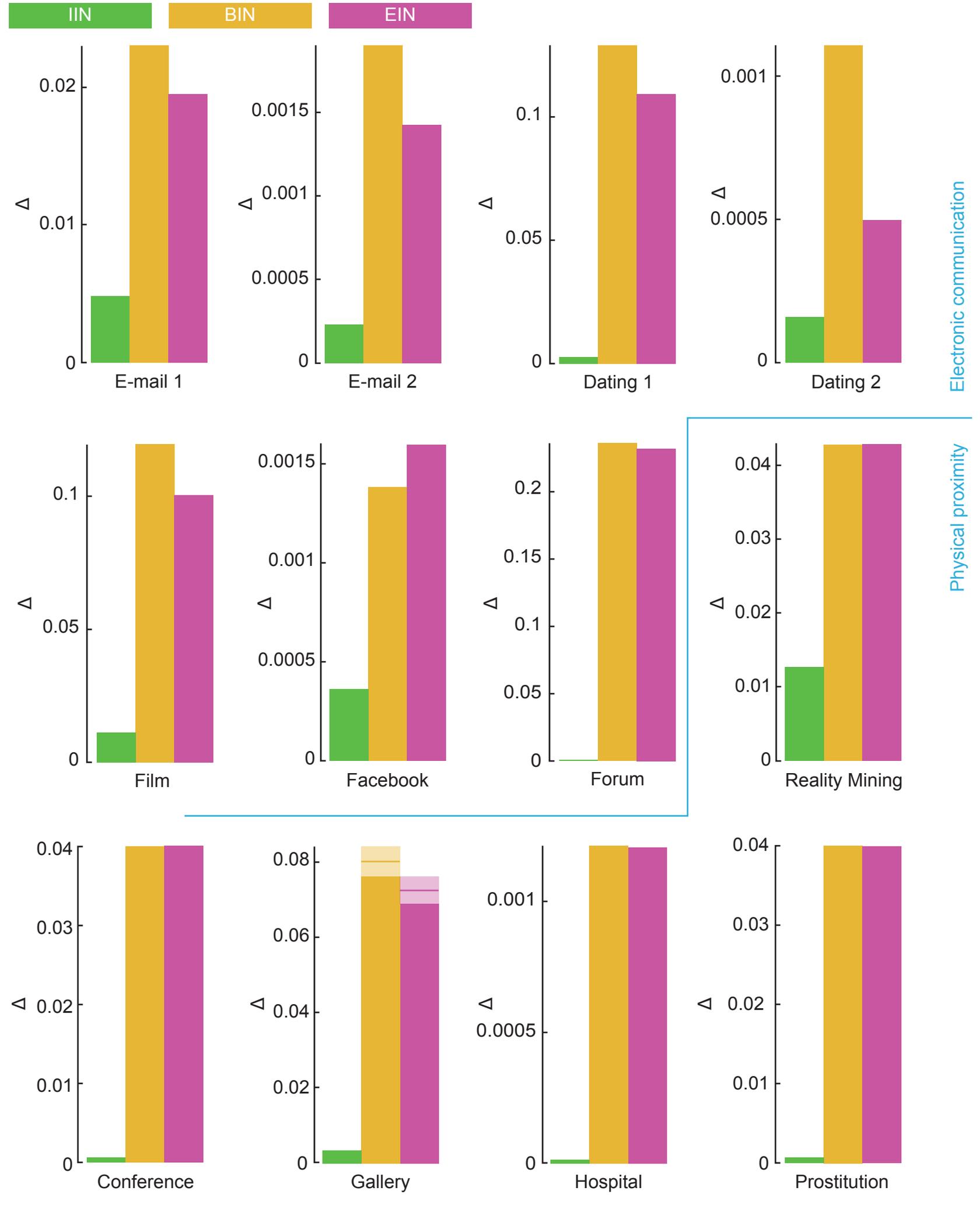

|  | Number of individuals | Number of contacts | Sampling time | Time resolution |
|---|---:|---:|---:|---:|
| E-mail 1 | 57,189 | 444,160 | 112.0d | 1s |
| E-mail 2 | 3,188 | 115,684 | 81.6d | 1s |
| Dating 1 | 28,972 | 529,890 | 512.0d | 1s |
| Dating 2 | 80,682 | 4,337,203 | 63.7d | 1s |
| Film | 35,624 | 472,496 | 8.27y | 1s |
| Facebook | 293,878 | 876,993 | 4.36y | 1s |
| Forum | 7,084 | 1,412,401 | 8.61y | 1s |
| Reality mining | 64 | 26,260 | 8.63h | 5s |
| Conference | 113 | 20,818 | 2.5d | 20s |
| Gallery | 159(8) | 6,027(350) | 7.3(1)h | 20s |
| Hospital | 293,878 | 64,625,283 | 9.77y | 1d |
| Prostitution | 16,730 | 50,632 | 6.00y | 1d |